\documentclass[pra,twocolumn]{revtex4-1}%
\usepackage{amsfonts}
\usepackage{amsmath}
\usepackage{amssymb}
\usepackage{graphicx}
\usepackage{hyperref}%
\providecommand{\U}[1]{\protect\rule{.1in}{.1in}}

\begin{document}
\preprint{ }
\title[ ]{Optimal estimation and discrimination of excess noise\\in thermal and amplifier channels}

\author{Masahiro Takeoka}
\affiliation{National Institute of Information and Communications Technology, 4-2-1
Nukui-kita, Koganei, Tokyo 184-8795, Japan}
\author{Mark M. Wilde}
\affiliation{Hearne Institute for Theoretical Physics, Department of Physics and Astronomy,
Center for Computation and Technology, Louisiana State University, Baton
Rouge, Louisiana 70803, USA}

\begin{abstract}
We determine a fundamental upper bound on the performance of any adaptive
protocol for discrimination or estimation of a channel which has an unknown
parameter encoded in the state of its environment. Since our approach relies
on the principle of data processing, the bound applies to a variety of
discrimination measures, including quantum relative entropy, hypothesis
testing relative entropy, R\'enyi relative entropy, fidelity, and quantum
Fisher information. We apply the upper bound to thermal (amplifier)\ channels
with a known transmissivity (gain)\ but unknown excess noise. In these cases,
we find that the upper bounds are achievable for several discrimination
measures of interest, and the method for doing so is non-adaptive, employing a
highly squeezed two-mode vacuum state at the input of each channel use.
Estimating the excess noise of a thermal channel is of principal interest for
the security of quantum key distribution, in the setting where a fiber-optic
cable has a known transmissivity but a tampering eavesdropper alters the
excess noise on the channel, so that estimating the excess noise as precisely
as possible is desirable. Finally, we outline a practical strategy which can
be used to achieve these limits.

\end{abstract}
\volumeyear{ }
\volumenumber{ }
\issuenumber{ }
\eid{ }
\date{\today}
\startpage{1}
\endpage{10}
\maketitle

\textit{Introduction}---One of the primary goals of quantum information theory
is to identify limitations on how well one can process information or estimate
an unknown parameter, when allowing for quantum effects
\cite{book2000mikeandike,H06book,H12,W15book}. Along with this goal, there is
great interest in determining whether it is possible to approach these limits
in principle, and furthermore, if this can be done in practice with realistic
constraints taken into account, such as time, energy, scalability, etc.

In this paper, we are interested in the fundamental limitations on channel
discrimination and estimation for a particular class of quantum channels.
Suppose that an unknown parameter $x$ is encoded in an environmental state,
which subsequently interacts with an input quantum system~$A$ via a fixed
unitary quantum interaction. Suppose further that the unitary interaction has
two output quantum systems, one of which is available and denoted as~$B$ and
the other is lost or discarded to the environment. The transformation of the
input system $A$ to the output system $B$ is called a quantum channel. Let us
call such channels \textit{environment-parametrized channels}, given that the
unknown parameter $x$\ is encoded exclusively in the environment and not in
the unitary interaction \footnote{These channels were called programmable
quantum channels in \cite{DP05,JWDFY08}, which is terminology used for them in
the context of quantum computation, the idea being that one could encode a
program in a quantum state that could then be executed via a unitary
interaction between an input and the program register. This meaning and
context is completely different from ours, so we prefer to use the terminology
``environment-parametrized channel''}. Important environment-parametrized
channels of practical interest are thermal channels with a fixed, known
transmissivity and unknown excess noise. Other examples are amplifier channels
with a fixed, known gain but unknown excess noise.

We consider two tasks:\ first, we suppose that the parameter $x$\ takes one of
two values and the goal is to figure out which value it takes. Second, we
suppose that the parameter $x$\ takes a value from a continuum and the goal is
to estimate the unknown parameter. The former task is called channel
discrimination
\cite{PhysRevA.71.062340,PhysRevA.72.014305,W08,DFY09,H09,HHLW10}\ and the
latter channel estimation \cite{CPR00,W02,FI03,JWDFY08,DKG12}, both topics
having an extensive literature already. Also, there are strong connections
between the two tasks \cite{N05}, as one might suspect. In these tasks, we
would like for the error probability or the mean-square error, respectively,
to be as small as possible when determining the unknown parameter.

\begin{figure}
[ptb]
\begin{center}
\includegraphics[
width=3.3901in
]%
{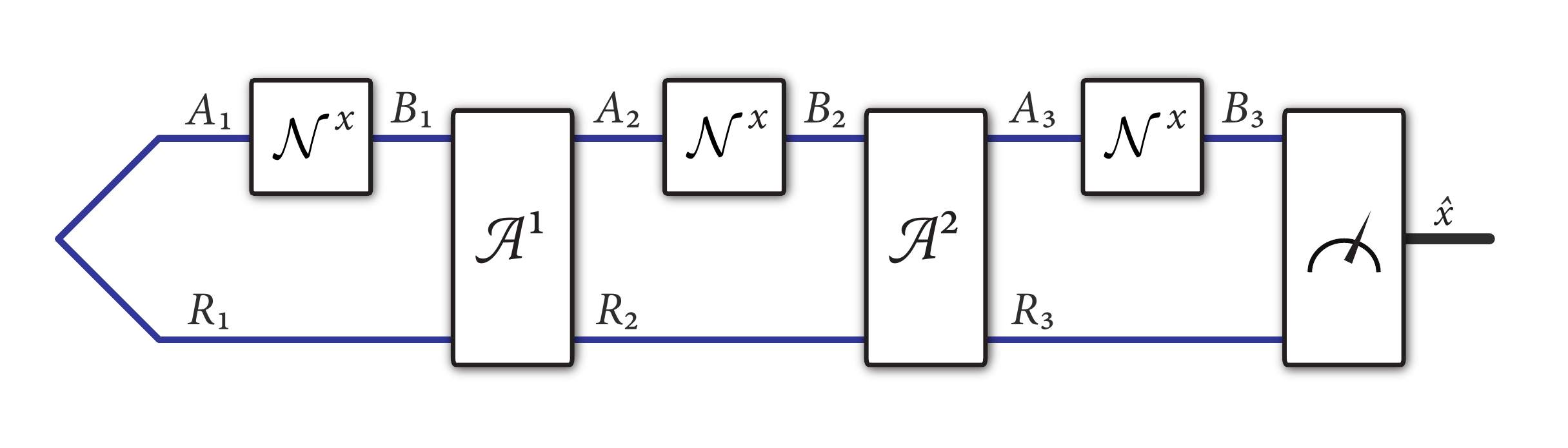}%
\caption{A depiction of the most general adaptive protocol for discriminating
or estimating quantum channels. A description of this protocol is given in the
main text.}%
\label{fig:adaptive-protocol}%
\end{center}
\end{figure}

For both tasks, the most general strategy one could allow for is an adaptive
strategy, when trying to determine an unknown parameter $x$ encoded in a
quantum channel $\mathcal{N}_{A\rightarrow B}^{x}$ (see
Figure~\ref{fig:adaptive-protocol}). An adaptive strategy that makes $M$ calls
to the channel is specified in terms of an input quantum state $\rho
_{R_{1}A_{1}}$, a set of adaptive, interleaved channels $\{\mathcal{A}%
_{R_{i}B_{i}\rightarrow R_{i+1}A_{i+1}}^{i}\}_{i=1}^{M-1}$, and a final
quantum measurement $\{\Lambda_{R_{M}B_{M}}^{\hat{x}}\}_{\hat{x}}$ that
outputs an estimate $\hat{x}$ of the unknown parameter. The strategy begins
with the discriminator preparing the input quantum state $\rho_{R_{1}A_{1}}$
and sending the $A_{1}$ system into the channel $\mathcal{N}_{A_{1}\rightarrow
B_{1}}^{x}$. The channel $\mathcal{N}_{A_{1}\rightarrow B_{1}}^{x}$ outputs
the system $B_{1}$, which is then available to the discriminator. The
discriminator adjoins the system $B_{1}$ to system $R_{1}$ and applies the
channel $\mathcal{A}_{R_{1}B_{1}\rightarrow R_{2}A_{2}}^{1}$. We say that the
channel $\mathcal{A}_{R_{1}B_{1}\rightarrow R_{2}A_{2}}^{1}$ is adaptive
because it can take an action conditioned on information in the system $B_{1}%
$, which itself might contain some partial information about the unknown
parameter~$x$. The discriminator then inputs the system $A_{2}$ into the
second use of the channel $\mathcal{N}_{A_{2}\rightarrow B_{2}}^{x}$, which
outputs a system $B_{2}$. This process repeats $M-2$ more times, and at the
end, the discriminator has systems $R_{M}$ and $B_{M}$. The discriminator
finally performs a measurement $\{\Lambda_{R_{M}B_{M}}^{\hat{x}}\}_{\hat{x}}$
that outputs an estimate $\hat{x}$ of the unknown parameter $x$. The
conditional probability for the estimate $\hat{x}$\ given the unknown
parameter $x$ is given by the Born rule:%
\begin{multline}
p_{\hat{X}|X}(\hat{x}|x)=\label{eq:cond-prob-adaptive}\\
\operatorname{Tr}\{\Lambda_{R_{M}B_{M}}^{\hat{x}}(\mathcal{N}_{A_{M}%
\rightarrow B_{M}}^{x}\circ\mathcal{A}_{R_{M-1}B_{M-1}\rightarrow R_{M}A_{M}%
}^{M-1}\circ\\
\cdots\circ\mathcal{A}_{R_{1}B_{1}\rightarrow R_{2}A_{2}}^{1}\circ
\mathcal{N}_{A_{1}\rightarrow B_{1}}^{x})(\rho_{R_{1}A_{1}})\}
\end{multline}
Note that such an adaptive strategy contains a non-adaptive strategy as a
special case: the system $R_{1}$ can be arbitrarily large and divided into
subsystems, with the only role of the interleaved channels $\mathcal{A}%
_{R_{i}B_{i}\rightarrow R_{i+1}A_{i+1}}^{i}$ being that they redirect these
subsystems to be the inputs of future calls to the channel (as would be the
case in any non-adaptive strategy for estimation or discrimination).%

Our first main result is a general upper bound on the performance of adaptive
discrimination and estimation of environment-parametrized channels. We
establish this upper bound for any discrimination measure that satisfies a
data-processing inequality (that is, it is monotone non-increasing with
respect to the action of a quantum channel). Our result thus holds for all
known and useful discrimination measures, given that the data-processing
inequality is the most basic requirement needed for any discrimination
measure. This includes well known discrimination measures such as quantum
relative entropy \cite{U62}, R\'{e}nyi relative entropy
\cite{P86,MDSFT13,WWY13}, quantum fidelity \cite{U76}, trace distance,
Chernoff information \cite{PhysRevLett.98.160501,ANSV08}, hypothesis testing
relative entropy \cite{HP91,BD10,WR12}, etc., each of which have operational
interpretations for certain information-processing tasks. The essential
statement of the upper bound is that one's ability to discriminate or estimate
environment-parametrized channels is limited by how well one can discriminate
or estimate the environmental states that encode the unknown parameter.

In our second main result, we show that it is possible to attain this upper
bound in principle for a number of the discrimination measures listed above,
when estimating excess noise in thermal channels or excess noise in amplifier
channels. For these particular channels, the unknown parameter is the mean
photon number of an environmental thermal state, while the transmissivity or
gain is known in our scenario. We find that the optimal strategy does not
involve any adaptation whatsoever and consists solely in sending one share of
a highly squeezed two-mode squeezed vacuum state into each use of the channel,
followed by a measurement on the output systems. What we find remarkable about
this result is that, in the limit of large squeezing, several of the
discrimination measures mentioned above depend only on the mean photon number
of the environmental thermal state and have no dependence on the
transmissivity or gain of the channel. Thus, such a strategy with a highly
squeezed two-mode squeezed vacuum state allows for removing the effect of loss
or gain in the channel, and we provide a physical interpretation for this
phenomenon in what follows.

Our results for estimating excess noise in thermal channels should be useful
for the security of quantum key distribution \cite{SBCDLP09}. There, the
transmissivity is typically known when the communication medium is a
fiber-optic cable, but the excess noise in the channel can be attributed to a
tampering eavesdropper. Thus, estimating excess noise in the channel is of
primary interest and plays a critical role in security analyses.

\textit{Environment-parametrized channels}---We begin by defining an
environment-parametrized quantum channel \cite{DP05,JWDFY08}. Let $x$ be an
unknown parameter, and let $\theta_{E}^{x}$ be a quantum state that depends on
$x$. Let $U_{AE\rightarrow BE^{\prime}}$ be a unitary operator that takes
vectors in a tensor-product input Hilbert space $\mathcal{H}_{A}%
\otimes\mathcal{H}_{E}$ to vectors in a tensor-product output Hilbert space
$\mathcal{H}_{B}\otimes\mathcal{H}_{E^{\prime}}$. Then we define an
environment-parametrized channel $\mathcal{N}_{A\rightarrow B}^{x}$\ as
follows:%
\begin{equation}
\mathcal{N}_{A\rightarrow B}^{x}(L_{A})\equiv\operatorname{Tr}_{E^{\prime}%
}\{U_{AE\rightarrow BE^{\prime}}(L_{A}\otimes\theta_{E}^{x})(U_{AE\rightarrow
BE^{\prime}})^{\dag}\}, \label{eq:env-param-ch}%
\end{equation}
where $L_{A}$ is an operator acting on $\mathcal{H}_{A}$ and
$\operatorname{Tr}_{E^{\prime}}$ denotes the partial trace. By inspecting the
above definition, we see that it is only the environment state $\theta_{E}%
^{x}$ that depends on the unknown parameter $x$ and the unitary interaction
$U_{AE\rightarrow BE^{\prime}}$ is fixed and independent of $x$. Thus, all of
the information that distinguishes one channel $\mathcal{N}_{A\rightarrow
B}^{x_{1}}$ from another channel $\mathcal{N}_{A\rightarrow B}^{x_{2}}$ is
encoded in the environment of these channels.

Particular examples of environment-parametrized channels are thermal channels,
noisy amplifier channels, Pauli channels, and erasure channels. We review the
first two here and sketch later why the latter two are
environment-parametrized. The unitary $U_{AE\rightarrow BE^{\prime}}$ for a
thermal channel is defined from the following Heisenberg input-output
relations:%
\begin{align}
\hat{b}  &  =\sqrt{\eta}\hat{a}+\sqrt{1-\eta}\hat{e}%
,\label{eq:thermal-channel}\\
\hat{e}^{\prime}  &  =-\sqrt{1-\eta}\hat{a}+\sqrt{\eta}\hat{e},
\label{eq:amplifier-channel}%
\end{align}
where $\hat{a}$, $\hat{b}$, $\hat{e}$, and $\hat{e}^{\prime}$ are the
field-mode annihilation operators for the sender's input, the receiver's
output, the environment's input, and the environment's output of these
channels, respectively. The environmental mode $\hat{e}$\ is prepared in a
thermal state $\theta(N_{B})$\ of mean photon number $N_{B}\geq0$, defined as%
\begin{equation}
\theta(N_{B})\equiv\frac{1}{N_{B}+1}\sum_{n=0}^{\infty}\left(  \frac{N_{B}%
}{N_{B}+1}\right)  ^{n}|n\rangle\langle n|, \label{eq:thermal-state}%
\end{equation}
where $\left\{  |n\rangle\right\}  $ is the orthonormal, photonic number-state
basis. The parameter $N_{B}$ is the excess noise of the thermal channel. When
$N_{B}=0$, $\theta(N_{B})$ reduces to the vacuum state, in which case the
resulting channel in \eqref{eq:thermal-channel} is called the pure-loss
channel---it is said to be quantum-limited in this case because the
environment is injecting the minimum amount of noise allowed by quantum
mechanics. The parameter $\eta\in\left[  0,1\right]  $ is the transmissivity
of the channel, representing the average fraction of photons making it from
the input to the output of the channel. Let $\mathcal{L}_{\eta,N_{B}}$ denote
this channel. In our application, we set the unknown parameter $x=N_{B}$, and
we suppose that the transmissivity $\eta$ is known.

The unitary $U_{AE\rightarrow BE^{\prime}}$ for an amplifier channel is
defined from the following Heisenberg input-output relations:
\begin{align}
\hat{b}  &  =\sqrt{G}\hat{a}+\sqrt{G-1}\hat{e}^{\dag},\\
\hat{e}^{\prime\dag}  &  =\sqrt{G-1}\hat{a}+\sqrt{G}\hat{e}^{\dag}.
\end{align}
The parameter $G\geq1$ is the gain of the amplifier channel. For this channel,
the environment is prepared in the thermal state $\theta(N_{B})$. The
parameter $N_{B}$ is the excess noise of the amplifier channel. If $N_{B}=0$,
the amplifier channel is said to be quantum-limited for a similar reason as
stated above. Let $\mathcal{A}_{G,N_{B}}$ denote this channel. The class of
amplifier channels we consider are those with a fixed known gain $G$ and the
unknown parameter $x=N_{B}$.

\textit{General bound from quantum data processing}---We now establish our
first main result. Let $\mathbf{D}(\rho\Vert\sigma)$ denote a generalized
divergence \cite{SW12,WWY13}, which is a function accepting two quantum states
as input and producing a non-negative real number as its output. The only
property that we demand to hold for a generalized divergence is that the
following data-processing inequality hold:%
\begin{equation}
\mathbf{D}(\rho\Vert\sigma)\geq\mathbf{D}(\mathcal{N}(\rho)\Vert
\mathcal{N}(\sigma)), \label{eq:GD-data-proc}%
\end{equation}
where $\mathcal{N}$ is a quantum channel. The inequality in
\eqref{eq:GD-data-proc} asserts that a generalized divergence $\mathbf{D}$,
interpreted as a measure of distinguishability of the states $\rho$ and
$\sigma$, does not increase under the action of a quantum channel
$\mathcal{N}$. Particular examples of generalized divergences include quantum
relative entropy \cite{U62}, hypothesis testing relative entropy
\cite{HP91,BD10,WR12}, quantum fidelity \cite{U76}, trace distance, R\'{e}nyi
relative entropy \cite{P86,MDSFT13,WWY13}, etc. Note that any generalized
divergence is unitarily invariant \cite{WWY13}: i.e., the following equality
holds for any unitary operator $U$:%
\begin{equation}
\mathbf{D}(\rho\Vert\sigma)=\mathbf{D}(U\rho U^{\dag}\Vert U\sigma U^{\dag}),
\label{eq:U-inv}%
\end{equation}
because $(\cdot)\rightarrow U(\cdot)U^{\dag}$ and $(\cdot)\rightarrow U^{\dag
}(\cdot)U$ are quantum channels, $\mathbf{D}(\rho\Vert\sigma)\geq
\mathbf{D}(U\rho U^{\dag}\Vert U\sigma U^{\dag})$, and $\mathbf{D}(U\rho
U^{\dag}\Vert U\sigma U^{\dag})\geq\mathbf{D}(U^{\dag}[U\rho U^{\dag}]U\Vert
U^{\dag}[U\sigma U^{\dag}]U)=\mathbf{D}(\rho\Vert\sigma)$. Furthermore, it is
invariant with respect to tensoring in the same state $\tau$ \cite{WWY13}:%
\begin{equation}
\mathbf{D}(\rho\Vert\sigma)=\mathbf{D}(\rho\otimes\tau\Vert\sigma\otimes\tau),
\label{eq:tens-inv}%
\end{equation}
because $(\cdot)\rightarrow(\cdot)\otimes\tau$ is a quantum channel and
partial trace is a quantum channel, so that $\mathbf{D}(\rho\Vert\sigma
)\geq\mathbf{D}(\rho\otimes\tau\Vert\sigma\otimes\tau)$ and $\mathbf{D}%
(\rho\otimes\tau\Vert\sigma\otimes\tau)\geq\mathbf{D}(\rho\Vert\sigma)$.

Suppose that the discriminator is attempting to distinguish two
environment-parametrized channels of the form in \eqref{eq:env-param-ch},
where the environmental state is either $\theta_{E}^{x_{1}}$ or $\theta
_{E}^{x_{2}}$. In such a case, the conditional probability for outputting
$\hat{x}$ is $p_{\hat{X}|X}(\hat{x}|x_{i})$ for $i\in\{1,2\}$ as given in
\eqref{eq:cond-prob-adaptive}, whenever the discrimination strategy is the
most general adaptive strategy as outlined before. Then our first main result
is the following inequality%
\begin{equation}
\mathbf{D}([\theta_{E}^{x_{1}}]^{\otimes M}\Vert\lbrack\theta_{E}^{x_{2}%
}]^{\otimes M})\geq\mathbf{D}(p_{\hat{X}|X}(\hat{x}|x_{1})\Vert p_{\hat{X}%
|X}(\hat{x}|x_{2})). \label{eq:1st-main-result}%
\end{equation}
Manifest in the above inequality is the following intuitive statement:\ the
discriminator's ability to distinguish the two channels, if given $M$ calls to
the channel, cannot be any better than if the discriminator were presented
with $M$ copies of the environmental state $\theta_{E}^{x_{i}}$ and then asked
to decide with which one he was presented. If the generalized divergence is
also additive with respect to tensor-product states, which holds for many
examples of divergences as we discuss below, then \eqref{eq:1st-main-result}
reduces to%
\begin{equation}
M\mathbf{D}(\theta_{E}^{x_{1}}\Vert\theta_{E}^{x_{2}})\geq\mathbf{D}%
(p_{\hat{X}|X}(\hat{x}|x_{1})\Vert p_{\hat{X}|X}(\hat{x}|x_{2})).
\label{eq:1-main-result-additive}%
\end{equation}
We note that results bearing some similarities to \eqref{eq:1st-main-result}
have appeared in previous papers \cite{JWDFY08,DKG12}, but the previous
statements are not given in such generality (i.e., for all generalized
divergences)\ nor were the previous statements argued to apply to the most
general adaptive strategy one could consider and instead only argued for
non-adaptive strategies.

We now prove the inequality in \eqref{eq:1st-main-result}. For simplicity, let
us suppose that the adaptive discrimination strategy consists of two calls to
the unknown channel, and then it will be easy to see how to generalize the
result to get \eqref{eq:1st-main-result}. Then, in this case,%
\begin{multline}
p_{\hat{X}|X}(\hat{x}|x_{i})=\operatorname{Tr}\{\Lambda_{R_{2}B_{2}}^{\hat{x}%
}(\mathcal{N}_{A_{2}\rightarrow B_{2}}^{x_{i}}\circ\mathcal{A}_{R_{1}%
B_{1}\rightarrow R_{2}A_{2}}^{1}\\
\circ\mathcal{N}_{A_{1}\rightarrow B_{1}}^{x_{i}})(\rho_{R_{1}A_{1}})\},
\end{multline}
and let us abbreviate the expression on the right as $\operatorname{Tr}%
\{\Lambda^{\hat{x}}\mathcal{N}^{x_{i}}\mathcal{A}^1\mathcal{N}^{x_{i}}(\rho)\}$. Then%
\begin{align}
&  \mathbf{D}(p_{\hat{X}|X}(\hat{x}|x_{1})\Vert p_{\hat{X}|X}(\hat{x}%
|x_{2}))\nonumber\\
&  \leq\mathbf{D}(\mathcal{N}^{x_{1}}\mathcal{A}^{1}\mathcal{N}^{x_{1}}%
(\rho)\Vert\mathcal{N}^{x_{2}}\mathcal{A}^{1}\mathcal{N}^{x_{2}}%
(\rho))\nonumber\\
&  \leq\mathbf{D}(U(\mathcal{A}^{1}\mathcal{N}^{x_{1}}(\rho)\otimes\theta
_{E}^{x_{1}})U^{\dag}\Vert U(\mathcal{A}^{1}\mathcal{N}^{x_{2}}(\rho
)\otimes\theta_{E}^{x_{2}})U^{\dag})\nonumber\\
&  =\mathbf{D}(\mathcal{A}^{1}\mathcal{N}^{x_{1}}(\rho)\otimes\theta
_{E}^{x_{1}}\Vert\mathcal{A}^{1}\mathcal{N}^{x_{2}}(\rho)\otimes\theta
_{E}^{x_{2}})\nonumber\\
&  \leq\mathbf{D}(\mathcal{N}^{x_{1}}(\rho)\otimes\theta_{E}^{x_{1}}%
\Vert\mathcal{N}^{x_{2}}(\rho)\otimes\theta_{E}^{x_{2}})\nonumber\\
&  \leq\mathbf{D}(U(\rho\otimes\theta_{E}^{x_{1}})U^{\dag}\otimes\theta
_{E}^{x_{1}}\Vert U(\rho\otimes\theta_{E}^{x_{2}})U^{\dag}\otimes\theta
_{E}^{x_{2}})\nonumber\\
&  =\mathbf{D}(\rho\otimes\theta_{E}^{x_{1}}\otimes\theta_{E}^{x_{1}}\Vert
\rho\otimes\theta_{E}^{x_{2}}\otimes\theta_{E}^{x_{2}})\nonumber\\
&  =\mathbf{D}(\theta_{E}^{x_{1}}\otimes\theta_{E}^{x_{1}}\Vert\theta
_{E}^{x_{2}}\otimes\theta_{E}^{x_{2}}).\label{eq:dp-method}%
\end{align}
All of the steps given above are a consequence of the data-processing
inequality in \eqref{eq:GD-data-proc}. The first inequality follows because
the final measurement can be considered as a quantum channel acting on the
states $\mathcal{N}^{x_{1}}\mathcal{A}^{1}\mathcal{N}^{x_{1}}(\rho)$ and
$\mathcal{N}^{x_{2}}\mathcal{A}^{1}\mathcal{N}^{x_{2}}(\rho)$ that produces
the respective output probability distributions $p_{\hat{X}|X}(\hat{x}|x_{1})$
and $p_{\hat{X}|X}(\hat{x}|x_{2})$. The second inequality follows from the
definition of environment-parametrized channels in
\eqref{eq:env-param-ch}\ and because a partial trace is a quantum channel. The
first equality follows because any generalized divergence is unitarily
invariant, as recalled in \eqref{eq:U-inv}. The third inequality follows by
discarding the adaptive channel $\mathcal{A}^{1}$. The next few steps follow
the same reasoning and the last equality follows from \eqref{eq:tens-inv}.
Thus we establish the inequality in \eqref{eq:1st-main-result} for $M=2$, but
it is easy to see that repeating the above steps establishes
\eqref{eq:1st-main-result} for arbitrary $M$.

\textit{Examples of generalized divergences}---One notable generalized
divergence is the quantum hypothesis testing relative entropy $D_{H}%
^{\varepsilon}(\rho\Vert\sigma)$ \cite{HP91,BD10,WR12}, defined for
$\varepsilon\in\left[  0,1\right]  $ as follows:%
\begin{equation}
D_{H}^{\varepsilon}(\rho\Vert\sigma)\equiv-\log\inf_{\Lambda}\operatorname{Tr}%
\{\Lambda\sigma\},
\end{equation}
where the infimum is with respect to all operators $\Lambda$ satisfying
$0\leq\Lambda\leq I$ and $\operatorname{Tr}\{\Lambda\rho\}\geq1-\varepsilon$.
The physical interpretation of this quantity is in asymmetric hypothesis
testing: if it is desired that the error probability in identifying the state
$\rho$ by a measurement $\{\Lambda,I-\Lambda\}$ be less than $\varepsilon$,
then $\inf_{\Lambda}\operatorname{Tr}\{\Lambda\sigma\}$ is the minimum error
that one could have in identifying the state $\sigma$ using the same
binary-outcome quantum measurement. The hypothesis testing relative entropy is
a quantity of deep interest in quantum information theory because various
relevant information measures can be built from it, which are useful in
assessing the performance of a variety of information-processing tasks
\cite{MW12,TH12,DTW14,TWW14,WTB16}. It obeys the data processing inequality in
\eqref{eq:GD-data-proc}\ by its very definition, for the simple reason that
applying the same quantum channel to the states $\rho$ and $\sigma$ never
decreases the two different error probabilities discussed above \cite{WR12}.

Applying the result in \eqref{eq:1st-main-result} leads to the following
bound:%
\begin{multline}
D_{H}^{\varepsilon}(p_{\hat{X}|X}(\hat{x}|x_{1})\Vert p_{\hat{X}|X}(\hat
{x}|x_{2}))\leq D_{H}^{\varepsilon}([\theta_{E}^{x_{1}}]^{\otimes M}%
\Vert\lbrack\theta_{E}^{x_{2}}]^{\otimes M})\label{eq:hypo-test-expand}\\
=MD(\theta_{E}^{x_{1}}\Vert\theta_{E}^{x_{2}})+\sqrt{MV(\theta_{E}^{x_{1}%
}\Vert\theta_{E}^{x_{2}})}\Phi^{-1}(\varepsilon)+O(\log M),
\end{multline}
where, in the last equality, we have used the quantum relative entropy
$D(\rho\Vert\sigma)\equiv\operatorname{Tr}\{\rho\lbrack\log\rho-\log\sigma]\}$
\cite{U62}, the quantum relative entropy variance $V(\rho\Vert\sigma
)\equiv\operatorname{Tr}\{\rho\lbrack\log\rho-\log\sigma-D(\rho\Vert
\sigma)]^{2}\}$ \cite{li12,TH12}, the inverse of the cumulative Gaussian
distribution function $\Phi$, and an expansion of the hypothesis testing
relative entropy that holds for tensor-power states \cite{li12,TH12,DPR15}.
The bound in \eqref{eq:hypo-test-expand} thus places a fundamental limitation
on the performance of any adaptive channel discrimination strategy in the
context of asymmetric hypothesis testing.

Notable additive generalized divergences are given by the R\'{e}nyi relative
entropies \cite{P86,MDSFT13,WWY13}, defined for $\alpha\in(0,1)\cup(1,\infty)$
as%
\begin{align}
D_{\alpha}(\rho\Vert\sigma) &  \equiv\frac{1}{\alpha-1}\log\operatorname{Tr}%
\{\rho^{\alpha}\sigma^{1-\alpha}\},\\
\widetilde{D}_{\alpha}(\rho\Vert\sigma) &  \equiv\frac{2\alpha}{\alpha-1}%
\log\left\Vert \rho^{1/2}\sigma^{(1-\alpha)/2\alpha}\right\Vert _{2\alpha
},\label{eq:sandwiched-Renyi}%
\end{align}
where $\left\Vert A\right\Vert _{p}\equiv\lbrack\operatorname{Tr}\{\left\vert
A\right\vert ^{p}\}]^{1/p}$ and $\left\vert A\right\vert \equiv\sqrt{A^{\dag
}A}$. The first one $D_{\alpha}(\rho\Vert\sigma)$ satisfies
\eqref{eq:GD-data-proc} for $\alpha\in\lbrack0,1)\cup(1,2]$ \cite{P86}, and
the second one satisfies \eqref{eq:GD-data-proc} for $\alpha\in\lbrack
1/2,1)\cup(1,\infty]$ \cite{MDSFT13,WWY13,B13monotone,FL13,MO15}. Both are
additive with respect to tensor-product states, converge to the quantum
relative entropy in the limit as $\alpha\rightarrow1$, and thus satisfy
\eqref{eq:GD-data-proc} in this limit. Applying \eqref{eq:1st-main-result} we
find that%
\begin{align}
D_{\alpha}(p_{\hat{X}|X}(\hat{x}|x_{1})\Vert p_{\hat{X}|X}(\hat{x}|x_{2})) &
\leq MD_{\alpha}(\theta_{E}^{x_{1}}\Vert\theta_{E}^{x_{2}}),\nonumber\\
\widetilde{D}_{\alpha}(p_{\hat{X}|X}(\hat{x}|x_{1})\Vert p_{\hat{X}|X}(\hat
{x}|x_{2})) &  \leq M\widetilde{D}_{\alpha}(\theta_{E}^{x_{1}}\Vert\theta
_{E}^{x_{2}}),
\end{align}
for the ranges of $\alpha$ for which data processing holds. As these
quantities have operational meaning in the context of asymmetric hypothesis
testing as error exponents and strong converse exponents in the quantum
Hoeffding bound \cite{MO15}, the above inequalities place fundamental
limitations on the exponential convergence rate of error probabilities of
adaptive channel discrimination strategies in this setting (see also
\cite{CMW14}\ for results on adaptive channel discrimination and R\'{e}nyi relative
entropies).

A special case of the R\'{e}nyi relative entropy in
\eqref{eq:sandwiched-Renyi} when $\alpha=1/2$ is $-\log F(\rho,\sigma)$, the
negative logarithm \ of the quantum fidelity, the latter defined as
$F(\rho,\sigma)\equiv\left\Vert \sqrt{\rho}\sqrt{\sigma}\right\Vert _{1}^{2}$.
Applying \eqref{eq:1-main-result-additive}, we find that%
\begin{equation}
\label{eq:fid-adap-bound}
- M\log F(\theta_{E}^{x_{1}},\theta_{E}^{x_{2}})\geq 
-\log F(p_{\hat{X}|X}(\hat{x}|x_{1}),p_{\hat{X}|X}(\hat
{x}|x_{2})).
\end{equation}

An important measure in quantum estimation theory is the quantum Fisher
information \cite{Hel76,H82,BC94,BCM96}, related to quantum fidelity and
defined for a continuously parametrized set $\{\sigma^{x}\}_{x}$\ of states as \cite[Theorem~6.3]{H06book}%
\begin{align}
I_{F}(x;\{\sigma^{x}\}_{x})& \equiv\lim_{\delta\rightarrow0}8\left[  1-\sqrt
{F}(\sigma^{x},\sigma^{x+\delta})\right]  /\delta^{2}  \nonumber \\
&  = \lim_{\delta\rightarrow0}[-4\log
{F}(\sigma^{x},\sigma^{x+\delta})]  /\delta^{2}.
\label{eq:Fisher-info}
\end{align}
(See the appendix for a derivation of the second equality.)
The importance of the quantum Fisher information is that it is a lower bound
on the variance of an unbiased estimator $\hat{x}$ of $x$
\cite{Hel76,H82,BC94,BCM96}:%
\begin{equation}
\text{Var}(\hat{x}-x)\geq\left[  I_{F}(x;\{\sigma^{x}\}_{x})\right]  ^{-1}.
\end{equation}
One can apply the same reasoning to adaptive protocols for estimating an
unknown parameter $x$ encoded in a family $\{\mathcal{N}^{x}\}_{x}$ of
channels, and we find that%
\begin{equation}
\text{Var}(\hat{x}-x)\geq\left[  I_{F}^{(M)}(x;\{\mathcal{N}^{x}%
\}_{x})\right]  ^{-1},
\end{equation}
where $I_{F}^{(M)}(x;\{\mathcal{N}^{x}%
\}_{x})$ is the Fisher information with respect to the
conditional probability defined in \eqref{eq:cond-prob-adaptive}. Applying the
bound in \eqref{eq:fid-adap-bound} and the relation between fidelity and
Fisher information in \eqref{eq:Fisher-info}, we find that the following lower
bound holds when trying to estimate an unknown parameter $x$ encoded in a
family $\{\mathcal{N}^{x}\}_{x}$ of environment-parametrized channels of the
form in \eqref{eq:env-param-ch}:%
\begin{equation}
\text{Var}(\hat{x}-x)\geq\left[  MI_{F}(x;\{\theta_{E}^{x}\}_{x})\right]  ^{-1}.
\label{eq:adaptive-fisher}%
\end{equation}
We note that this inequality generalizes those from \cite{JWDFY08,DKG12,KD13},
given that those works did not consider adaptive protocols for estimating $x$.

\textit{Application to thermal channels}---We now show that several of the
above upper bounds are in fact achievable, whenever the goal is to determine
the excess noise in a thermal channel with known transmissivity. We begin with
channel discrimination. Suppose that we are given two thermal channels
$\mathcal{L}_{\eta,N_{B}^{1}}$ and $\mathcal{L}_{\eta,N_{B}^{2}}$, each having
a known transmissivity $\eta\in\lbrack0,1)$ with excess noise equal to
$N_{B}^{1}\geq0$ or $N_{B}^{2}\geq0$. (If $\eta=1$ or $N_{B}^{1}=N_{B}^{2}$,
then it is impossible to distinguish the channels and so we do not consider
these cases.)\ In all cases for discrimination or estimation, we find that a
non-adaptive strategy involving $M$ copies of a highly squeezed, two-mode
squeezed vacuum state suffices to attain the upper bounds given above, proving
that this non-adaptive strategy suffices for achieving the best possible
performance. The two-mode squeezed vacuum state is equivalent to a
purification of the thermal state in \eqref{eq:thermal-state}\ and is defined
as%
\begin{multline}
|\psi_{\operatorname{TMS}}(N_{S})\rangle_{RA}\\
\equiv\frac{1}{\sqrt{N_{S}+1}}\sum_{n=0}^{\infty}\sqrt{\left(  \frac{N_{S}%
}{N_{S}+1}\right)  ^{n}}|n\rangle_{R}|n\rangle_{A}.
\end{multline}
The strategy we are employing in all cases leads to the following, final
pre-measurement state for $i\in\{1,2\}$:%
\begin{equation}
\sigma_{N_{B}^{i}}\equiv\left[  (\operatorname{id}_{R}\otimes\mathcal{L}%
_{\eta,N_{B}^{i}})(|\psi_{\operatorname{TMS}}(N_{S})\rangle\langle
\psi_{\operatorname{TMS}}(N_{S})|_{RA})\right]  ^{\otimes M}.
\label{eq:thermal-channel-on-TMSV}%
\end{equation}

Starting with quantum relative entropy, we find the following expansion for
large $N_{S}$ and for $\eta\in\lbrack0,1)$, by employing a formula for the
quantum relative entropy of Gaussian states \cite{PhysRevA.71.062320,PLOB15}:%
\begin{align}
&  D(\sigma_{N_{B}^{1}}\Vert\sigma_{N_{B}^{2}})\nonumber\\
&  =-g(N_{B}^{1},N_{B}^{1})+g(N_{B}^{1},N_{B}^{2})+O(1/N_{S})\\
&  =D(\theta(N_{B}^{1})\Vert\theta(N_{B}^{2}))+O(1/N_{S}).
\label{eq:rel-ent-thermal-converg}%
\end{align}
where $g(x,y)$ is a relative entropic generalization of the well known formula
for the entropy of a bosonic thermal state (see, e.g., \cite{G08thesis}) and
is defined for $x,y\geq0$ as%
\begin{equation}
g(x,y)\equiv(x+1)\log(y+1)-x\log y.
\end{equation}
In fact, as indicated in \eqref{eq:rel-ent-thermal-converg}, we find for all
$\eta\in\lbrack0,1)$ that $\lim_{N_{S}\rightarrow\infty}D(\sigma_{N_{B}^{1}%
}\Vert\sigma_{N_{B}^{2}})=D(\theta(N_{B}^{1})\Vert\theta(N_{B}^{2}))$, so that
the relative entropy in the limit of high squeezing converges to the classical
relative entropy between the two thermal states that distinguish the channels
(here we say classical relative entropy because the states $\theta(N_{B}^{1})$
and $\theta(N_{B}^{2})$ commute).

Similarly, we find the following expansion for the quantum relative entropy
variance for large $N_{S}$ and for $\eta\in\lbrack0,1)$, by employing a
formula for the quantum relative entropy variance of Gaussian states
\cite{WTLB16}:%
\begin{align}
&  V(\sigma_{N_{B}^{1}}\Vert\sigma_{N_{B}^{2}})\nonumber\\
&  =N_{B}^{1}(N_{B}^{1}+1)\log^{2}\!\left(  \frac{1+1/N_{B}^{1}}{1+1/N_{B}%
^{2}}\right)  +O(1/N_{S})\label{eq:rel-ent-var-thermal}\\
&  =V(\theta(N_{B}^{1})\Vert\theta(N_{B}^{2}))+O(1/N_{S}).
\label{eq:rel-ent-var-converge}%
\end{align}
As indicated in \eqref{eq:rel-ent-var-converge}, we also find for all $\eta
\in\lbrack0,1)$ that $\lim_{N_{S}\rightarrow\infty}V(\sigma_{N_{B}^{1}}%
\Vert\sigma_{N_{B}^{2}})=V(\theta(N_{B}^{1})\Vert\theta(N_{B}^{2}))$. The
formula in \eqref{eq:rel-ent-var-thermal} is an expression for the relative
entropy variance of two thermal states, which generalizes the entropy variance
formula from \cite{WRG15}\ for a thermal state. See the appendix for a derivation.

By the statement in \eqref{eq:hypo-test-expand}, we find the following upper
bound on the performance of any adaptive strategy when discriminating the
channels
\begin{multline}
D_{H}^{\varepsilon}(p_{\hat{X}|X}(\hat{x}|N_{B}^{1})\Vert p_{\hat{X}|X}%
(\hat{x}|N_{B}^{2}))\leq MD(\theta(N_{B}^{1})\Vert\theta(N_{B}^{2}%
))\label{eq:thermal-channel-achieve}\\
+\sqrt{MV(\theta(N_{B}^{1})\Vert\theta(N_{B}^{2}))}\Phi^{-1}(\varepsilon
)+O(\log M).
\end{multline}
Since we know from prior work \cite{li12,TH12,DPR15}\ the following lower
bound on the hypothesis testing relative entropy%
\begin{multline}
D_{H}^{\varepsilon}(\sigma_{N_{B}^{1}}\Vert\sigma_{N_{B}^{2}})\geq
MD(\sigma_{N_{B}^{1}}\Vert\sigma_{N_{B}^{2}})\\
+\sqrt{MV(\sigma_{N_{B}^{1}}\Vert\sigma_{N_{B}^{2}})}\Phi^{-1}(\varepsilon
)+O(\log M),
\end{multline}
the expansions for large $N_{S}$ in \eqref{eq:rel-ent-thermal-converg} and
\eqref{eq:rel-ent-var-converge} establish that the upper bound in
\eqref{eq:thermal-channel-achieve}\ is achievable in the limit as
$N_{S}\rightarrow\infty$. As a consequence, by using a highly squeezed state
as a probe and in the limit of high squeezing, it is as if the loss in the
channel has no effect on the transmitted state and one's ability to
distinguish the channels is as good as one's ability to distinguish the
environmental states $\theta(N_{B}^{1})$ and $\theta(N_{B}^{2})$, which
correspond to the excess noise in the channels. We offer an explanation for
this phenomenon later on.

Turning to the fidelity, we find similar results. Applying a formula for the
fidelity of two-mode Gaussian states \cite{MM12}, we find for $\eta\in
\lbrack0,1)$ that%
\begin{align}
&  F(\sigma_{N_{B}^{1}},\sigma_{N_{B}^{2}})\nonumber\\
&  =\left[  \sqrt{\left(  N_{B}^{1}+1\right)  \left(  N_{B}^{2}+1\right)
}-\sqrt{N_{B}^{1}N_{B}^{2}}\right]  ^{-2}+O(1/N_{S})\\
&  =F(\theta(N_{B}^{1}),\theta(N_{B}^{2}))+O(1/N_{S}). \label{eq:fid-converge}%
\end{align}
Consistent with our previous observations and as indicated in
\eqref{eq:fid-converge}, we also find for $\eta\in\lbrack0,1)$ that
$\lim_{N_{S}\rightarrow\infty}F(\sigma_{N_{B}^{1}},\sigma_{N_{B}^{2}%
})=F(\theta(N_{B}^{1}),\theta(N_{B}^{2}))$.

We finally consider the quantum Fisher information $I_{F}(N_{B};\{\sigma
_{N_{B}}\}_{N_{B}})$ as defined in \eqref{eq:Fisher-info}. Applying a formula
for the fidelity of two-mode Gaussian states \cite{MM12} and expanding about
small $\delta>0$ and large $N_{S}$, we find for $N_{B}^{1}=N_{B}$ and
$N_{B}^{2}=N_{B}+\delta$ that%
\begin{multline}
\sqrt{F}(\sigma_{N_{B}^{1}},\sigma_{N_{B}^{2}})=1-\frac{1-\eta\left[
N_{S}\left(  1-\eta\right)  \left(  2N_{B}+1\right)  \right]  ^{-1}}%
{8N_{B}\left(  N_{B}+1\right)  }\delta^{2}\\
+O(\delta^{3}/N_{S}^{2}).
\end{multline}
Thus, by applying \eqref{eq:Fisher-info}, we find that the quantum Fisher
information in the large $N_{S}$ limit is equal to%
\begin{equation}
\lim_{N_{S}\rightarrow\infty}I_{F}(N_{B};\{\sigma_{N_{B}}\}_{N_{B}})=\frac
{1}{N_{B}\left(  N_{B}+1\right)  }, \label{eq:fisher-info-thermal}%
\end{equation}
in agreement with \cite[Eq.~(63)]{GL14}. By applying the bound from
\eqref{eq:adaptive-fisher}, the fact that the quantum Fisher information of an
ensemble $\{\theta(N_{B})\}_{N_{B}}$ of thermal states is equal to $\left[
N_{B}\left(  N_{B}+1\right)  \right]  ^{-1}$, and the fact that the quantum
Fisher information is achievable in principle by a measurement
\cite{Hel76,H82,BC94,BCM96}, we can conclude that there exists a non-adaptive
strategy that achieves the ultimate precision possible in the limit of high
squeezing. Furthermore, the form of the quantum Fisher information in
\eqref{eq:fisher-info-thermal} has an intuitive form:\ the noisier the state,
the lower the Fisher information, and vice versa.

\textit{Concrete Discrimination Strategy}---All of the convergences of the
quantum discrimination measures to the discrimination of two thermal states
begs for an intuitive explanation. Here we give some explanation for this
phenomenon, by establishing a physical relation between a thermal state with
mean photon number $N_{B}$ and the state $\sigma_{N_{B}}$ defined in
\eqref{eq:thermal-channel-on-TMSV}, in the limit as $N_{S}\rightarrow\infty$.
At the same time, this explanation leads to a concrete discrimination strategy
consisting of applying the unitary transformation given below followed by photodetection.

The Wigner characteristic function covariance matrix for $\sigma_{N_{B}}$ in
\eqref{eq:thermal-channel-on-TMSV} is as follows (see, e.g., \cite{PLOB15}):%
\begin{equation}
V=\left[
\begin{array}
[c]{cccc}%
a & c & 0 & 0\\
c & b & 0 & 0\\
0 & 0 & a & -c\\
0 & 0 & -c & b
\end{array}
\right]  ,
\end{equation}
where%
\begin{align}
a  &  =\eta N_{S}+(1-\eta)N_{B}+1/2,\\
b  &  =N_{S}+1/2,\\
c  &  =\sqrt{\eta N_{S}(N_{S}+1)}.
\end{align}
Consider the following symplectic transformation:%
\begin{equation}
S=\left[
\begin{array}
[c]{cccc}%
\omega_{+} & -\omega_{-} & 0 & 0\\
-\omega_{-} & \omega_{+} & 0 & 0\\
0 & 0 & \omega_{+} & \omega_{-}\\
0 & 0 & \omega_{-} & \omega_{+}%
\end{array}
\right]  ,
\end{equation}
where%
\begin{align}
\omega_{+}  &  =\sqrt{\frac{1+N_{S}}{1+(1-\eta)N_{S}}},\\
\omega_{-}  &  =\sqrt{\frac{\eta N_{S}}{1+(1-\eta)N_{S}}}.
\end{align}
The symplectic transformation $S$ is independent of  $N_{B}$ 
and diagonalizes $V$ when $N_{B}=0$. 
Also, $S$ can be realized by a two-mode squeezer,
which corresponds to a unitary transformation acting on the tensor-product
Hilbert space of the two modes. Applying $S$ to $V$ with finite $N_{B}$, we
get%
\begin{equation}
SVS^{T}=\left[
\begin{array}
[c]{cccc}%
a_{s} & -c_{s} & 0 & 0\\
-c_{s} & b_{s} & 0 & 0\\
0 & 0 & a_{s} & c_{s}\\
0 & 0 & c_{s} & b_{S}%
\end{array}
\right]  ,
\end{equation}
where%
\begin{align}
a_{s}  &  =N_{B}+1/2+O(N_{S}^{-1}),\\
b_{s}  &  =(1-\eta)N_{S}+\eta N_{B}+1/2+O(N_{S}^{-1}),\\
c_{s}  &  =\sqrt{\eta}N_{B}+O(N_{S}^{-1}),
\end{align}
One can physically eliminate the off-diagonal terms by randomizing the two
modes (or just by simply treating them separately). Then in the limit as
$N_{S}\rightarrow\infty$, we find that the above state is equivalent to a
product of two thermal states with photon numbers $N_{B}$ and $(1-\eta
)N_{S}+\eta N_{B}$. So a concrete discrimination strategy consists in applying
the above unitary transformation to the output of each channel, tracing over
the second mode, and performing photodetection on the first mode, which is the
optimal measurement for distinguishing two thermal states.

\textit{Application to amplifier channels}---For quantum amplifier channels
with a fixed known gain but unknown excess noise, we find results similar to
the ones given above for thermal channels. The upper bound from
\eqref{eq:1st-main-result}\ results in a generalized divergence between two
thermal states. Also,\ the quantum relative entropy, the quantum relative
entropy variance, the fidelity, and the quantum Fisher information evaluated
for the state $\left[  (\operatorname{id}_{R}\otimes\mathcal{A}_{G,N_{B}^{i}%
})(|\psi_{\operatorname{TMS}}(N_{S})\rangle\langle\psi_{\operatorname{TMS}%
}(N_{S})|_{RA})\right]  ^{\otimes M}$ converge to the same expressions given
above in the limit of high squeezing, having no dependence on the gain of the
amplifier channel. There is a similar explanation for the convergences as
given above and a resulting concrete discrimination strategy in the limit of
high squeezing.

\textit{Teleportation method}---One can also arrive at our results for thermal
and amplifier channels in terms of a technique called teleportation simulation
\cite[Section~V]{BDSW96}. In \cite[Section~V]{BDSW96}, the authors showed how any protocol
consisting of adaptive operations interleaved between many independent uses of
the same channel can be reduced to a non-adaptive protocol if the channel is
simulable by teleportation. This method was reviewed recently in
\cite{PLOB15}\ and therein extended to continuous-variable bosonic channels and others
as well. Recently, the technique was also applied in the context of channel
discrimination and estimation of particular channels \cite{PL16}.

Briefly, the main idea of the teleportation method is to 1)\ replace every
channel in the protocol by its simulation with teleportation and 2) rearrange
all of the uses of the channel to the start of the protocol, such that all of
the adaptive operations act at the end of the protocol and the resulting
protocol no longer has the adaptive form. For the channels considered in
\cite{PL16} (limited to Pauli channels or erasure channels), the resulting
protocol is such that one feeds in $M$ shares of a maximally entangled state
to each channel use. Then a final measurement is
performed on this state to discriminate two channels in a given class.

In the examples that we consider here, including thermal channels of a fixed
transmissivity or amplifier channels of a fixed gain, we can instead use the
two-mode squeezed vacuum state and continuous-variable teleportation \cite{prl1998braunstein} to effect
the teleportation reduction discussed above. One critical aspect of the
problem setup is that the channels being discriminated or estimated have the
same transmissivity or gain, so that the teleportation correction operations
are independent of the particular channel being discriminated or estimated. In
order for the teleportation simulation to be perfect, it is necessary to
consider the limit of high squeezing, as we have done above, and the result is
to recover all of the convergences of quantum discrimination measures
discussed previously.

The teleportation simulation approach to understanding our results is
interesting, but we think that the data-processing method outlined in this
paper is simpler and more powerful when applicable. The data-processing method applies
independently of whether a channel is teleportation simulable, and
furthermore, we only need a generalized divergence for the argument in
\eqref{eq:dp-method} to hold, whereas one further requires continuity (albeit
a natural property) in order for the teleportation argument to go through in
the continuous-variable case. Finally, the data-processing method outlined
here recovers all of the results established in \cite{PL16} because all of the
channels considered there are in fact environment-parametrized. To see this,
for Pauli channels, we can take the environment state $\theta_{E}^{x}$ in
\eqref{eq:env-param-ch}\ to be $\sum_{i=0}^{d^{2}-1}p_{i}|i\rangle\langle
i|_{E}$ and the unitary interaction to be $\sum_{i=0}^{d^{2}-1}U_{A}%
^{i}\otimes|i\rangle\langle i|_{E}$, where the parameter $x$ is the
probability vector $\{p_{i}\}_{i}$ and $U_{A}^{i}$ is a Pauli operator. For
erasure channels, we can take the environment state in
\eqref{eq:env-param-ch}\ to be $\sum_{i=0}^{1}p_{i}|i\rangle\langle i|_{E_{1}%
}\otimes|e\rangle\langle e|_{E_{2}}$ and the unitary interaction to be
$|0\rangle\langle0|_{E_{1}}\otimes I_{AE_{2}}+|1\rangle\langle1|\otimes
\operatorname{SWAP}_{AE_{2}}$.

It would be interesting to determine if there are teleportation-simulable
channels that are not environment-parametrized. If it were the case, then the
teleportation simulation method could be used to analyze adaptive
discrimination and estimation protocols, whereas the data-processing method
would not necessarily apply.

\textit{Conclusion}---We have outlined a general method for bounding the
performance of adaptive channel discrimination or estimation of
environment-parametrized channels, in which an unknown parameter is encoded in
the environment of the channel. The method applies to any generalized
divergence, a function whose sole property is data processing (monotonicity
under the action of a quantum channel). We applied the approach to several
discrimination measures that have operational meaning in a variety of
contexts. As a concrete example, we considered thermal (amplifier)\ channels
with known transmissivity (gain) and unknown excess noise. We derived
limitations on the performance of the most general adaptive discrimination or
estimation strategies for these channels, and we also showed that these limits
are achievable in principle if highly squeezed states are available.

Going forward from here, it would be interesting to generalize the approach to
channels encoding multiple unknown parameters that need to be estimated or
discriminated---the results from \cite{GL14,M13}\ should be helpful here, at
least in the case of quantum Gaussian channels. We also wonder whether there
are other approaches, besides the data-processing method or the teleportation
simulation approach, that could be used to simplify adaptive protocols for
channel discrimination or estimation.

\textit{Acknowledgements}---We are grateful to Saikat Guha and Chenglong You
for discussions related to the topic of this paper. MT is grateful to the
Hearne Institute for Theoretical Physics at Louisiana State University for
hosting him during October 2016, when this research was completed. MMW
acknowledges support from the NSF under Award No.~CCF-1350397.

\textit{Appendix}---Here we establish the formula in
\eqref{eq:rel-ent-var-thermal} for the relative entropy variance of two
thermal states and the formula in \eqref{eq:Fisher-info} for the quantum Fisher information.

We begin by establishing \eqref{eq:rel-ent-var-thermal}. Let%
\begin{align}
\rho &  =\frac{1}{N_{B}^{1}+1}\sum_{n=0}^{\infty}\left(  \frac{N_{B}^{1}%
}{N_{B}^{1}+1}\right)  ^{n}|n\rangle\langle n|,\\
\sigma &  =\frac{1}{N_{B}^{2}+1}\sum_{n=0}^{\infty}\left(  \frac{N_{B}^{2}%
}{N_{B}^{2}+1}\right)  ^{n}|n\rangle\langle n|.
\end{align}
The relative entropy variance is defined as%
\begin{equation}
V(\rho\Vert\sigma)=\operatorname{Tr}\{\rho\left[  \log\rho-\log\sigma
-D(\rho\Vert\sigma)\right]  ^{2}\}.
\end{equation}
Consider that%
\begin{align}
D(\rho\Vert\sigma)  &  =-g(N_{B}^{1},N_{B}^{1})+g(N_{B}^{1},N_{B}^{2})\\
&  =-(N_{B}^{1}+1)\log(N_{B}^{1}+1)+N_{B}^{1}\log N_{B}^{1}\nonumber\\
&  \qquad+(N_{B}^{1}+1)\log(N_{B}^{2}+1)-N_{B}^{1}\log N_{B}^{2}.
\end{align}
Also,%
\begin{align}
&  \log\rho-\log\sigma\nonumber\\
&  =-\log(N_{B}^{1}+1)+\log(N_{B}^{2}+1)\nonumber\\
&  \qquad+\sum_{n=0}^{\infty}n\log\left(  \frac{N_{B}^{1}}{N_{B}^{1}+1}\left[
\frac{N_{B}^{2}}{N_{B}^{2}+1}\right]  ^{-1}\right)  |n\rangle\langle n|\\
&  =-\log(N_{B}^{1}+1)+\log(N_{B}^{2}+1)\nonumber\\
&  \qquad+\hat{n}\log\left(  \frac{N_{B}^{1}}{N_{B}^{1}+1}\left[  \frac
{N_{B}^{2}}{N_{B}^{2}+1}\right]  ^{-1}\right),
\end{align}
where $\hat{n}$ is the photon-number operator.
So then%
\begin{align}
&  \log\rho-\log\sigma-D(\rho\Vert\sigma)\nonumber\\
&  =-N_{B}^{1}\log(N_{B}^{1}+1)+N_{B}^{1}\log N_{B}^{1}\nonumber\\
&  \qquad+N_{B}^{1}\log(N_{B}^{2}+1)-N_{B}^{1}\log N_{B}^{2}\nonumber\\
&  \qquad+\hat{n}\log\left(  \frac{N_{B}^{1}}{N_{B}^{1}+1}\left[  \frac
{N_{B}^{2}}{N_{B}^{2}+1}\right]  ^{-1}\right) \\
&  =\left(  \hat{n}-N_{B}^{1}\right)  \log\left(  \frac{N_{B}^{1}}{N_{B}%
^{1}+1}\left[  \frac{N_{B}^{2}}{N_{B}^{2}+1}\right]  ^{-1}\right),
\end{align}
and%
\begin{multline}
\left[  \log\rho-\log\sigma-D(\rho\Vert\sigma)\right]  ^{2}\\
=\left(  \hat{n}-N_{B}^{1}\right)  ^{2}\log^{2}\left(  \frac{N_{B}^{1}}%
{N_{B}^{1}+1}\left[  \frac{N_{B}^{2}}{N_{B}^{2}+1}\right]  ^{-1}\right).
\end{multline}
Finally, we find that%
\begin{align}
&  V(\rho\Vert\sigma)\nonumber\\
&  =\left\langle \left(  \hat{n}-N_{B}^{1}\right)  ^{2}\right\rangle
_{\theta(N_{B}^{1})}\log^{2}\left(  \frac{N_{B}^{1}}{N_{B}^{1}+1}\left[
\frac{N_{B}^{2}}{N_{B}^{2}+1}\right]  ^{-1}\right) \\
&  =N_{B}^{1}(N_{B}^{1}+1)\log^{2}\left(  \frac{N_{B}^{1}}{N_{B}^{1}+1}\left[
\frac{N_{B}^{2}}{N_{B}^{2}+1}\right]  ^{-1}\right).
\end{align}

\bigskip
Now we derive the formula in \eqref{eq:Fisher-info} for the quantum Fisher information:%
\begin{equation}
I_{F}(x;\{\rho_{x}\}_{x})=\lim_{\delta\rightarrow0}\left[  -4\log F(\rho
_{x},\rho_{x+\delta})\right]  /\delta^{2}.\label{eq:fisher-log-2nd-deriv}%
\end{equation}
To begin with, consider the known formula for quantum Fisher information \cite[Theorem~6.3]{H06book}:%
\begin{equation}
\lim_{\delta\rightarrow0}\frac{8\left[  1-\sqrt{F}(\rho_{x},\rho_{x+\delta
})\right]  }{\delta^{2}}.
\end{equation}
To evaluate this, we apply L'Hospital's rule, and find that%
\begin{align}
&  \lim_{\delta\rightarrow0}-8\frac{\frac{d}{d\delta}\sqrt{F}(\rho_{x}%
,\rho_{x+\delta})}{2\delta}\nonumber\\
&  =\lim_{\delta\rightarrow0}\left[  -4\frac{d^{2}}{d\delta^{2}}\sqrt{F}%
(\rho_{x},\rho_{x+\delta})\right]  \\
&  =-4\left.  \frac{d^{2}}{d\delta^{2}}\sqrt{F}(\rho_{x},\rho_{x+\delta
})\right\vert _{\delta=0},
\end{align}
so that%
\begin{equation}
I_{F}(x;\{\rho_{x}\}_{x})=-4\left.  \frac{d^{2}}{d\delta^{2}}\sqrt{F}(\rho
_{x},\rho_{x+\delta})\right\vert _{\delta=0}.\label{eq:rewrite-Fisher-info}%
\end{equation}

Now we move on to showing \eqref{eq:fisher-log-2nd-deriv}. Consider that%
\begin{equation}
-2\log F(\rho_{x},\rho_{x+\delta})=-4\log\sqrt{F}(\rho_{x},\rho_{x+\delta}).
\end{equation}
Furthermore,%
\begin{align}
&  \frac{d^{2}}{d\delta^{2}}\left[  -\log\sqrt{F}(\rho_{x},\rho_{x+\delta
})\right]  \nonumber\\
&  =\frac{d}{d\delta}\left[  \frac{d}{d\delta}\left[  -\log\sqrt{F}(\rho
_{x},\rho_{x+\delta})\right]  \right]  \\
&  =\frac{d}{d\delta}\left[  -\left(  \sqrt{F}(\rho_{x},\rho_{x+\delta
})\right)  ^{-1}\left(  \frac{d}{d\delta}\sqrt{F}(\rho_{x},\rho_{x+\delta
})\right)  \right]  \\
&  =\left(  \sqrt{F}(\rho_{x},\rho_{x+\delta})\right)  ^{-2}\left(  \frac
{d}{d\delta}\sqrt{F}(\rho_{x},\rho_{x+\delta})\right)  ^{2}\nonumber\\
&  \qquad-\left(  \sqrt{F}(\rho_{x},\rho_{x+\delta})\right)  ^{-1}\left(
\frac{d^{2}}{d\delta^{2}}\sqrt{F}(\rho_{x},\rho_{x+\delta})\right)  .
\end{align}
Then we find that%
\begin{align}
&  \left.  \frac{d^{2}}{d\delta^{2}}\left[  -\log\sqrt{F}(\rho_{x}%
,\rho_{x+\delta})\right]  \right\vert _{\delta=0}\nonumber\\
&  =\left.  \left(  \sqrt{F}(\rho_{x},\rho_{x+\delta})\right)  ^{-2}\left(
\frac{d}{d\delta}\sqrt{F}(\rho_{x},\rho_{x+\delta})\right)  ^{2}\right\vert
_{\delta=0}\nonumber\\
&  \qquad-\left.  \left(  \sqrt{F}(\rho_{x},\rho_{x+\delta})\right)
^{-1}\left(  \frac{d^{2}}{d\delta^{2}}\sqrt{F}(\rho_{x},\rho_{x+\delta
})\right)  \right\vert _{\delta=0}\\
&  =\left.  \left(  \frac{d}{d\delta}\sqrt{F}(\rho_{x},\rho_{x+\delta
})\right)  ^{2}\right\vert _{\delta=0}-\left.  \frac{d^{2}}{d\delta^{2}}%
\sqrt{F}(\rho_{x},\rho_{x+\delta})\right\vert _{\delta=0}.
\end{align}
The quantity%
\begin{equation}
\left.  \left(  \frac{d}{d\delta}\sqrt{F}(\rho_{x},\rho_{x+\delta})\right)
^{2}\right\vert _{\delta=0}=0
\end{equation}
because%
\begin{align}
& \frac{d}{d\delta}\sqrt{F}(\rho_{x},\rho_{x+\delta}) \nonumber\\
&  =\frac{d}{d\delta
}\operatorname{Tr}\left\{  \sqrt{\sqrt{\rho_{x}}\rho_{x+\delta}\sqrt{\rho_{x}%
}}\right\}  \\
&  =\frac{1}{2}\operatorname{Tr}\left\{  \left(  \sqrt{\rho_{x}}\rho
_{x+\delta}\sqrt{\rho_{x}}\right)  ^{-1/2}\sqrt{\rho_{x}}\frac{d}{d\delta}%
(\rho_{x+\delta})\sqrt{\rho_{x}}\right\}  ,
\end{align}
and so%
\begin{align}
&  \left.  \frac{d}{d\delta}\sqrt{F}(\rho_{x},\rho_{x+\delta})\right\vert
_{\delta=0}\nonumber\\
&  =\frac{1}{2}\operatorname{Tr}\left\{  \left(  \sqrt{\rho_{x}}\rho_{x}%
\sqrt{\rho_{x}}\right)  ^{-1/2}\sqrt{\rho_{x}}\left(  \left.  \frac{d}%
{d\delta}\rho_{x+\delta}\right\vert _{\delta=0}\right)  \sqrt{\rho_{x}%
}\right\}  \\
&  =\frac{1}{2}\operatorname{Tr}\left\{  \rho_{x}^{-1}\sqrt{\rho_{x}}\left(
\left.  \frac{d}{d\delta}\rho_{x+\delta}\right\vert _{\delta=0}\right)
\sqrt{\rho_{x}}\right\}  \\
&  =\frac{1}{2}\operatorname{Tr}\left\{  \left.  \frac{d}{d\delta}%
\rho_{x+\delta}\right\vert _{\delta=0}\right\}  \\
&  =0,
\end{align}
where the last line follows from the definition of the derivative and because
the difference of two density operators is equal to zero. (In the above we
assumed that the density operators $\rho_{x}$ are full rank but one can arrive
at the same conclusion when they are not necessarily full rank \cite{Liu2014167}.) So we
find that%
\begin{multline}
\left.  \frac{d^{2}}{d\delta^{2}}\left[  -2\log F(\rho_{x},\rho_{x+\delta
})\right]  \right\vert _{\delta=0}\\
=-4\left.  \frac{d^{2}}{d\delta^{2}}\sqrt{F}(\rho_{x},\rho_{x+\delta
})\right\vert _{\delta=0},
\end{multline}
which is consistent with \eqref{eq:rewrite-Fisher-info}.

Thus we can conclude \eqref{eq:fisher-log-2nd-deriv} because after applying
L'Hospital's rule, we find that%
\begin{align}
&  \lim_{\delta\rightarrow0}\frac{-4\log F(\rho_{x},\rho_{x+\delta})}%
{\delta^{2}}\nonumber\\
&  =\lim_{\delta\rightarrow0}\frac{-\frac{d}{d\delta}4\log F(\rho_{x}%
,\rho_{x+\delta})}{2\delta}\\
&  =\lim_{\delta\rightarrow0}\left[  -\frac{d^{2}}{d\delta^{2}}2\log
F(\rho_{x},\rho_{x+\delta})\right]  .
\end{align}

\bibliographystyle{unsrt}
\bibliography{Ref}

\begin{thebibliography}{10}

\bibitem{book2000mikeandike}
Michael~A. Nielsen and Isaac~L. Chuang.
\newblock {\em Quantum Computation and Quantum Information}.
\newblock Cambridge University Press, 2000.

\bibitem{H06book}
Masahito Hayashi.
\newblock {\em Quantum Information: An Introduction}.
\newblock Springer, 2006.

\bibitem{H12}
Alexander~S. Holevo.
\newblock {\em Quantum Systems, Channels, Information}.
\newblock de Gruyter Studies in Mathematical Physics (Book 16). de Gruyter,
  November 2012.

\bibitem{W15book}
Mark~M. Wilde.
\newblock {\em From Classical to Quantum Shannon Theory}.
\newblock March 2016.
\newblock arXiv:1106.1445v7.

\bibitem{Note1}
These channels were called programmable quantum channels in \cite
  {DP05,JWDFY08}, which is terminology used for them in the context of quantum
  computation, the idea being that one could encode a program in a quantum
  state that could then be executed via a unitary interaction between an input
  and the program register. This meaning and context is completely different
  from ours, so we prefer to use the terminology ``environment-parametrized
  channel''.

\bibitem{PhysRevA.71.062340}
Massimiliano~F. Sacchi.
\newblock Optimal discrimination of quantum operations.
\newblock {\em Physical Review A}, 71(6):062340, June 2005.
\newblock arXiv:quant-ph/0505183.

\bibitem{PhysRevA.72.014305}
Massimiliano~F. Sacchi.
\newblock Entanglement can enhance the distinguishability of
  entanglement-breaking channels.
\newblock {\em Physical Review A}, 72(1):014305, July 2005.
\newblock arXiv:quant-ph/0505174.

\bibitem{W08}
John Watrous.
\newblock Distinguishing quantum operations having few {Kraus} operators.
\newblock {\em Quantum Information and Computation}, 8(9):819--833, 2008.
\newblock arXiv:0710.0902.

\bibitem{DFY09}
Runyao Duan, Yuan Feng, and Mingsheng Ying.
\newblock Perfect distinguishability of quantum operations.
\newblock {\em Physical Review Letters}, 103(21):210501, November 2009.
\newblock arXiv:0908.0119.

\bibitem{H09}
Masahito Hayashi.
\newblock Discrimination of two channels by adaptive methods and its
  application to quantum system.
\newblock {\em IEEE Transactions on Information Theory}, 55(8):3807--3820,
  August 2009.
\newblock arXiv:0804.0686.

\bibitem{HHLW10}
Aram~W. Harrow, Avinatan Hassidim, Debbie Leung, and John Watrous.
\newblock Adaptive versus non-adaptive strategies for quantum channel
  discrimination.
\newblock {\em Physical Review A}, 81(3):032339, March 2010.
\newblock arXiv:0909.0256.

\bibitem{CPR00}
Andrew~M. Childs, John Preskill, and Joseph Renes.
\newblock Quantum information and precision measurement.
\newblock {\em Journal of Modern Optics}, 47(2-3):155--176, 2000.
\newblock arXiv:quant-ph/9904021.

\bibitem{W02}
Howard~M. Wiseman.
\newblock Adaptive quantum-limited estimates of phase.
\newblock {\em Australian Optical Society News}, 16:14--19, 2002.
\newblock arXiv:quant-ph/0206124.

\bibitem{FI03}
Akio Fujiwara and Hiroshi Imai.
\newblock Quantum parameter estimation of a generalized {Pauli} channel.
\newblock {\em Journal of Physics A: Mathematical and General}, 36(29):8093,
  July 2003.

\bibitem{JWDFY08}
Zhengfeng Ji, Guoming Wang, Runyao Duan, Yuan Feng, and Mingsheng Ying.
\newblock Parameter estimation of quantum channels.
\newblock {\em IEEE Transactions on Information Theory}, 54(11):5172--5185,
  November 2008.
\newblock arXiv:quant-ph/0610060.

\bibitem{DKG12}
Rafal Demkowicz-Dobrzanski, Jan Kolodynski, and Madalin Guta.
\newblock The elusive {Heisenberg} limit in quantum-enhanced metrology.
\newblock {\em Nature Communications}, 3:1063, September 2012.
\newblock arXiv:1201.3940.

\bibitem{N05}
Hiroshi Nagaoka.
\newblock {\em Asymptotic Theory of Quantum Statistical Inference}, chapter On
  the Relation between {Kullback} Divergence and {Fisher} Information: From
  Classical Systems to Quantum Systems, pages 399--419.
\newblock World Scientific, 2005.

\bibitem{U62}
Hisaharu Umegaki.
\newblock Conditional expectations in an operator algebra {IV} (entropy and
  information).
\newblock {\em Kodai Mathematical Seminar Reports}, 14(2):59--85, 1962.

\bibitem{P86}
D\'enes Petz.
\newblock Quasi-entropies for finite quantum systems.
\newblock {\em Reports in Mathematical Physics}, 23(1):57--65, February 1986.

\bibitem{MDSFT13}
Martin {M\"uller}-Lennert, Fr\'ed\'eric Dupuis, Oleg Szehr, Serge Fehr, and
  Marco Tomamichel.
\newblock On quantum {R\'enyi} entropies: a new generalization and some
  properties.
\newblock {\em Journal of Mathematical Physics}, 54(12):122203, December 2013.
\newblock arXiv:1306.3142.

\bibitem{WWY13}
Mark~M. Wilde, Andreas Winter, and Dong Yang.
\newblock Strong converse for the classical capacity of entanglement-breaking
  and {Hadamard} channels via a sandwiched {R\'enyi} relative entropy.
\newblock {\em Communications in Mathematical Physics}, 331(2):593--622,
  October 2014.
\newblock arXiv:1306.1586.

\bibitem{U76}
Armin Uhlmann.
\newblock The ``transition probability'' in the state space of a *-algebra.
\newblock {\em Reports on Mathematical Physics}, 9(2):273--279, 1976.

\bibitem{PhysRevLett.98.160501}
K.~M.~R. Audenaert, J.~Calsamiglia, R.~Mu\ noz Tapia, E.~Bagan, Ll. Masanes,
  A.~Acin, and F.~Verstraete.
\newblock Discriminating states: The quantum {Chernoff} bound.
\newblock {\em Physical Review Letters}, 98:160501, April 2007.
\newblock arXiv:quant-ph/0610027.

\bibitem{ANSV08}
K.~M.~R. Audenaert, M.~Nussbaum, A.~Szkola, and F.~Verstraete.
\newblock Asymptotic error rates in quantum hypothesis testing.
\newblock {\em Communications in Mathematical Physics}, 279:251--283, 2008.
\newblock arXiv:0708.4282.

\bibitem{HP91}
Fumio Hiai and D\'enes Petz.
\newblock The proper formula for relative entropy and its asymptotics in
  quantum probability.
\newblock {\em Communications in Mathematical Physics}, 143(1):99--114,
  December 1991.

\bibitem{BD10}
Francesco Buscemi and Nilanjana Datta.
\newblock The quantum capacity of channels with arbitrarily correlated noise.
\newblock {\em IEEE Transactions on Information Theory}, 56(3):1447--1460,
  March 2010.
\newblock arXiv:0902.0158.

\bibitem{WR12}
Ligong Wang and Renato Renner.
\newblock One-shot classical-quantum capacity and hypothesis testing.
\newblock {\em Physical Review Letters}, 108(20):200501, May 2012.
\newblock arXiv:1007.5456.

\bibitem{SBCDLP09}
Valerio Scarani, Helle Bechmann-Pasquinucci, Nicolas~J. Cerf, Miloslav
  Du\ifmmode~\check{s}\else \v{s}\fi{}ek, Norbert L\"utkenhaus, and Momtchil
  Peev.
\newblock The security of practical quantum key distribution.
\newblock {\em Reviews of Modern Physics}, 81(3):1301--1350, September 2009.
\newblock arXiv:0802.4155.

\bibitem{DP05}
Giacomo~Mauro D'Ariano and Paolo Perinotti.
\newblock Programmable quantum channels and measurements.
\newblock In {\em Workshop on Quantum Information Theory and Quantum
  Statistical Inference}, Tokyo, Japan. ERATO Quantum Computation and
  Information Project, November 2005.
\newblock arXiv:quant-ph/0510033.

\bibitem{SW12}
Naresh Sharma and Naqueeb~Ahmad Warsi.
\newblock On the strong converses for the quantum channel capacity theorems.
\newblock June 2012.
\newblock arXiv:1205.1712.

\bibitem{MW12}
William Matthews and Stephanie Wehner.
\newblock Finite blocklength converse bounds for quantum channels.
\newblock {\em IEEE Transactions on Information Theory}, 60(11):7317--7329,
  November 2014.
\newblock arXiv:1210.4722.

\bibitem{TH12}
Marco Tomamichel and Masahito Hayashi.
\newblock A hierarchy of information quantities for finite block length
  analysis of quantum tasks.
\newblock {\em IEEE Transactions on Information Theory}, 59(11):7693--7710,
  November 2013.
\newblock arXiv:1208.1478.

\bibitem{DTW14}
Nilanjana Datta, Marco Tomamichel, and Mark~M. Wilde.
\newblock On the second-order asymptotics for entanglement-assisted
  communication.
\newblock {\em Quantum Information Processing}, 15(6):2569--2591, June 2016.
\newblock arXiv:1405.1797.

\bibitem{TWW14}
Marco Tomamichel, Mark~M. Wilde, and Andreas Winter.
\newblock Strong converse rates for quantum communication.
\newblock June 2014.
\newblock arXiv:1406.2946.

\bibitem{WTB16}
Mark~M. Wilde, Marco Tomamichel, and Mario Berta.
\newblock Converse bounds for private communication over quantum channels.
\newblock February 2016.
\newblock arXiv:1602.08898.

\bibitem{li12}
Ke~Li.
\newblock Second order asymptotics for quantum hypothesis testing.
\newblock {\em Annals of Statistics}, 42(1):171--189, February 2014.
\newblock arXiv:1208.1400.

\bibitem{DPR15}
Nilanjana Datta, Yan Pautrat, and Cambyse Rouz\'{e}.
\newblock Second-order asymptotics for quantum hypothesis testing in settings
  beyond i.i.d. - quantum lattice systems and more.
\newblock {\em Journal of Mathematical Physics}, 57(6):062207, June 2016.
\newblock arXiv:1510.04682.

\bibitem{B13monotone}
Salman Beigi.
\newblock Sandwiched {R\'enyi} divergence satisfies data processing inequality.
\newblock {\em Journal of Mathematical Physics}, 54(12):122202, December 2013.
\newblock arXiv:1306.5920.

\bibitem{FL13}
Rupert~L. Frank and Elliott~H. Lieb.
\newblock Monotonicity of a relative {R\'enyi} entropy.
\newblock {\em Journal of Mathematical Physics}, 54(12):122201, December 2013.
\newblock arXiv:1306.5358.

\bibitem{MO15}
Mil\'an Mosonyi and Tomohiro Ogawa.
\newblock Quantum hypothesis testing and the operational interpretation of the
  quantum {R\'enyi} relative entropies.
\newblock {\em Communications in Mathematical Physics}, 334(3):1617--1648,
  March 2015.
\newblock arXiv:1309.3228.

\bibitem{CMW14}
Tom Cooney, Milan Mosonyi, and Mark~M. Wilde.
\newblock Strong converse exponents for a quantum channel discrimination
  problem and quantum-feedback-assisted communication.
\newblock {\em Communications in Mathematical Physics}, 344(3):797--829, June
  2016.
\newblock arXiv:1408.3373.

\bibitem{Hel76}
Carl~W. Helstrom.
\newblock {\em Quantum Detection and Estimation Theory}.
\newblock Academic, New York, 1976.

\bibitem{H82}
Alexander~S. Holevo.
\newblock {\em Probabilistic and statistical aspects of quantum theory}.
\newblock North-Holland, Amsterdam, 1982.

\bibitem{BC94}
Samuel~L. Braunstein and Carlton~M. Caves.
\newblock Statistical distance and the geometry of quantum states.
\newblock {\em Physical Review Letters}, 72(22):3439--3443, May 1994.

\bibitem{BCM96}
Samuel~L. Braunstein, Carlton~M. Caves, and Gerard~J. Milburn.
\newblock Generalized uncertainty relations: Theory, examples, and lorentz
  invariance.
\newblock {\em Annals of Physics}, 247:135--173, 1996.
\newblock arXiv:quant-ph/9507004.

\bibitem{KD13}
Jan Kolodynski and Rafal Demkowicz-Dobrzanski.
\newblock Efficient tools for quantum metrology with uncorrelated noise.
\newblock {\em New Journal of Physics}, 15(7):073043, July 2013.
\newblock arXiv:1303.7271.

\bibitem{PhysRevA.71.062320}
Xiao-yu Chen.
\newblock Gaussian relative entropy of entanglement.
\newblock {\em Physical Review A}, 71(6):062320, June 2005.
\newblock arXiv:quant-ph/0402109.

\bibitem{PLOB15}
Stefano Pirandola, Riccardo Laurenza, Carlo Ottaviani, and Leonardo Banchi.
\newblock 2015.
\newblock arXiv:1510.08863.

\bibitem{G08thesis}
Saikat Guha.
\newblock {\em Multiple-User Quantum Information Theory for Optical
  Communication Channels}.
\newblock PhD thesis, Massachusetts Institute of Technology, June 2008.

\bibitem{WTLB16}
Mark~M. Wilde, Marco Tomamichel, Seth Lloyd, and Mario Berta.
\newblock Gaussian hypothesis testing and quantum illumination.
\newblock August 2016.
\newblock arXiv:1608.06991.

\bibitem{WRG15}
Mark~M. Wilde, Joseph~M. Renes, and Saikat Guha.
\newblock Second-order coding rates for pure-loss bosonic channels.
\newblock {\em Quantum Information Processing}, 15(3):1289--1308, March 2016.
\newblock arXiv:1408.5328.

\bibitem{MM12}
Paulina Marian and Tudor~A. Marian.
\newblock Uhlmann fidelity between two-mode {Gaussian} states.
\newblock {\em Physical Review A}, 86(2):022340, August 2012.
\newblock arXiv:1111.7067.

\bibitem{GL14}
Yang Gao and Hwang Lee.
\newblock Bounds on quantum multiple-parameter estimation with {Gaussian}
  state.
\newblock {\em The European Physical Journal D}, 68(11):347, November 2014.
\newblock arXiv:1407.7352.

\bibitem{BDSW96}
Charles~H. Bennett, David~P. DiVincenzo, John~A. Smolin, and William~K.
  Wootters.
\newblock Mixed-state entanglement and quantum error correction.
\newblock {\em Physical Review A}, 54(5):3824--3851, November 1996.
\newblock arXiv:quant-ph/9604024.

\bibitem{PL16}
Stefano Pirandola and Cosmo Lupo.
\newblock September 2016.
\newblock arXiv:1609.02160.

\bibitem{prl1998braunstein}
Samuel~L. Braunstein and H.~J. Kimble.
\newblock Teleportation of continuous quantum variables.
\newblock {\em Physical Review Letters}, 80(4):869--872, January 1998.

\bibitem{M13}
Alex Monras.
\newblock Phase space formalism for quantum estimation of {Gaussian} states.
\newblock March 2013.
\newblock arXiv:1303.3682.

\bibitem{Liu2014167}
Jing Liu, Heng-Na Xiong, Fei Song, and Xiaoguang Wang.
\newblock Fidelity susceptibility and quantum fisher information for density
  operators with arbitrary ranks.
\newblock {\em Physica A: Statistical Mechanics and its Applications},
  410:167--173, September 2014.
\newblock arXiv:1401.3154.

\end{thebibliography}

\end{document}